\documentclass{WileyMSP-template}

\usepackage[square, numbers, comma, sort&compress,super]{natbib}
\usepackage{ragged2e}%

\usepackage[font={sf,bf},textfont=md]{caption}

\usepackage[colorlinks,
            linkcolor=blue,
            anchorcolor=blue,
            citecolor=blue,
            ]{hyperref}

\usepackage{color}
\usepackage{amsmath}
\usepackage{amsfonts}
\usepackage{amssymb}
\usepackage{graphicx}
\usepackage{tablefootnote}
\usepackage{threeparttable}
\usepackage{multirow}
\usepackage{textcomp}
\usepackage{float}
\usepackage{breakurl}
\usepackage{autobreak}

\begin{document}

\pagestyle{fancy}
\rhead{\includegraphics[width=2.5cm]{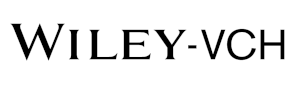}}

\title{Commensurate Stacking Phase Transitions in an Intercalated Transition Metal Dichalcogenide}

\maketitle


\author{Xiaohui Yang}
\author{Jin-Ke Bao}
\author{Zhefeng Lou}
\author{Peng Li}
\author{Chenxi Jiang}
\author{Jialu Wang}
\author{Tulai Sun}
\author{Yabin Liu}
\author{Wei Guo}
\author{Sitaram Ramakrishnan}
\author{Surya Rohith Kotla}
\author{Martin Tolkiehn}
\author{Carsten Paulmann}
\author{Guang-Han Cao}
\author{Yuefeng Nie}
\author{Wenbin Li}
\author{Yang Liu}
\author{Sander van Smaalen*}
\author{Xiao Lin*}
\author{Zhu-An Xu*}

\dedication{}


\begin{affiliations}

Dr. X. Yang, Z. Lou, Dr. P. Li, C. Jiang, J. Wang, Y. Liu, Prof. G.-H. Cao, Prof. Y. Liu, Prof. Z.-A. Xu\\
Zhejiang Province Key Laboratory of Quantum Technology and Device, Department of Physics, Zhejiang University, Hangzhou 310027, P. R. China\\
Email: zhuan@zju.edu.cn

\vspace{0.5em}
Dr. X. Yang, Z. Lou, J. Wang, Prof. X. Lin\\
Key Laboratory for Quantum Materials of Zhejiang Province, School of Science, Westlake University, Hangzhou 310024, P. R. China\\
Email: linxiao@westlake.edu.cn

\vspace{0.5em}
Dr. X. Yang, Z. Lou, J. Wang, Prof. X. Lin\\
Institute of Natural Sciences, Westlake Institute for Advanced Study, Hangzhou 310024, P. R. China\\

\vspace{0.5em}
Dr. J.-K. Bao, Dr. S. Ramakrishnan, S. R. Kotla, Prof. S. van Smaalen\\
Laboratory of Crystallography, University of Bayreuth, 95447 Bayreuth, Germany\\
Email: smash@uni-bayreuth.de

\vspace{0.5em}
Dr. J.-K. Bao\\
Department of Physics, Materials Genome Institute and International Center for Quantum and Molecular Structures, Shanghai University, Shanghai 200444, P. R. China\\

\vspace{0.5em}
Dr. P. Li, Prof. Y. Liu\\
Center for Correlated Matter, Zhejiang University, Hangzhou 310058, P. R. China\\

\vspace{0.5em}
Dr. T. Sun\\
Center for Electron Microscopy, State Key Laboratory Breeding Base of Green Chemistry Synthesis Technology and College of Chemical Engineering, Zhejiang University of Technology, Hangzhou 310014, P. R. China\\

\vspace{0.5em}
Dr. T. Sun\\
School of Materials Science and Engineering, Zhejiang University, Hangzhou 310027, P. R. China\\

\vspace{0.5em}
Dr. W. Guo, Prof. Y. Nie\\
National Laboratory of Solid State Microstructures, College of Engineering and Applied Sciences, and Collaborative Innovation Center of Advanced Microstructures, Nanjing University, Nanjing 210093, P. R. China\\

\vspace{0.5em}
Dr. S. Ramakrishnan\\
Department of Quantum Matter, AdSM, Hiroshima University, Higashi-Hiroshima 739-8530, Japan\\

\vspace{0.5em}
Dr. M. Tolkiehn\\
P24, PETRA III, DESY, 22607 Hamburg, Germany\\

\vspace{0.5em}
Dr. C. Paulmann\\
Mineralogisch-Petrographisches Institute, Universit\"{a}t Hamburg, 20146 Hamburg, Germany\\

\vspace{0.5em}
Prof. G.-H. Cao, Prof. Z.-A. Xu\\
State Key Lab of Silicon Materials, Zhejiang University, Hangzhou 310027, P. R. China\\
~\\
\vspace{0.5em}
Prof. W. Li\\
Key Laboratory of 3D Micro/Nano Fabrication and Characterization of Zhejiang Province, School of Engineering, Westlake University, Hangzhou 310024, Zhejiang Province, P. R. China\\

\end{affiliations}
~\\

\justifying  

\begin{abstract}
\justifying  
\noindent Intercalation and stacking order modulation are two active ways in manipulating the interlayer interaction of transition metal dichalcogenides (TMDCs), which lead to a variety of emergent phases and allow for engineering material properties. 
Herein, the growth of Pb intercalated TMDCs--Pb(Ta$_{1+x}$Se$_2$)$_2$, the first 124-phase, is reported. Pb(Ta$_{1+x}$Se$_2$)$_2$ exhibits a unique two-step first-order structural phase transition at around 230 K.
The transitions are solely associated with the stacking degree of freedom, evolving from a high temperature phase with ABC stacking and symmetry $R3m$ to an intermediate phase with AB stacking and $P3m1$, and finally to a low temperature phase with again symmetry $R3m$, but with ACB stacking. Each step involves a rigid slide of building blocks by a vector [1/3, 2/3, 0]. Intriguingly, gigantic lattice contractions occur at the transitions on warming. At low temperature, bulk superconductivity with $T_\textrm{c}\approx$ 1.8 K is observed. The underlying physics of the structural phase transitions are discussed from first-principle calculations. The symmetry analysis reveals topological nodal lines in the band structure. Our results demonstrate the possibility to realize higher order metal intercalated phases of TMDCs, advance our knowledge of polymorphic transitions and may inspire stacking order engineering in TMDCs and beyond.

\end{abstract}

\setlength{\parindent}{1em}

\section{Introduction}

\noindent The strong hierarchy between weak interlayer coupling and strong intralayer covalent bonding in TMDCs leads to multiple polymorphs with diverse stacking orders and transition metal M-atom coordinations sandwiched by chalcogens~\cite{Katzke2004}. 
Each phase has unique physical properties, which may include the formation of charge density wave (CDW)~\cite{Rossnagel2011}, Ising superconductivity~\cite{Mak2015,Mak2018}, valleytronics~\cite{Mak2012,Zeng2012}, quantum spin Hall insulator~\cite{Qian2014,Herrero2018}. Recently, the stacking degree of freedom became a crucial knob to manipulate a plethora of emergent phenomena, such as unconventional superconductivity, flat band dispersion and topological phase, as exemplified in magic-angle-twisted bilayer graphene~\cite{cao2018unsc} and Bernal~(ABA)/rhombohedral~(ABC)-stacking trilayer graphene~\cite{lui_trilayerc,bao_trilayerc}. For TMDCs, there is a fast-growing interest in this field: twisted bilayer TMDCs own much stronger correlations~\cite{zhang2020flat,wang2020} and bulk TMDCs with different stacking orders exhibit distinct emergent phenomena~\cite{Deng2016,Zhengjacs,MoS2valley,MoS2stacking}.

In addition to stacking-engineering, intercalation is another powerful tool to tune physical properties in TMDCs. On the other hand, intercalated TMDCs are important systems in the study of intercalation chemistry~\cite{wang_review}. The guest species can occupy diverse geometric voids such as trigonal prismatic or octahedral sites formed by the chalcogen atoms and even induce a stacking order transition in the original TMDCs~\cite{Brec1986}. 
The intercalation of additional species can either donate electrons, as in Cu$_{x}$TiSe$_{2}$~\cite{morosan2006SC} or distort original lattices~\cite{LixMoS22021,CuTaSe2_Sander,self_intercalated_Loh}. For instance,  MoS$_2$ undergoes a concomitant structural phase transition from semiconducting 2$H$-MoS$_2$ to metallic 1$T$-Li$_x$MoS$_2$ under electrochemical Li adsorption~\cite{LixMoS22021}. Pb intercalated PbTaSe$_{2}$ -- 112-phase~\cite{MTaX21980} exhibits topological superconductivity~\cite{Guan2016} with the absence of CDW transitions, in distinct contrast with that of 2$H$-TaSe$_{2}$~\cite{yokota2000superconductivity,CDW}.

Here, we explore more possibilities of the intercalation chemistry in the fertile playground of TMDCs. 
We synthesized 124-phase Pb(Ta$_{1+x}$Se$_{2}$)$_{2}$, a stage-2 intercalation compound~\cite{dresselhaus1981intercalation}, each two TaSe$_{2}$ layers separated by one Pb layer. To the best of our knowledge, such phase has not been reported in intercalated TMDCs before. More intriguingly, we observed two commensurate stacking phase transitions, evolving from high temperature (high-$T$) $R3m$ phase-ABC stacking to intermediate $P3m1$ phase-AB stacking, and to low temperature (low-$T$) ACB stacking phase but with the same space group as that of high-$T$ phase. Each step involves a rigid glide of [Se-Ta-Se-Ta$_{2x}$-Se-Ta-Se-Pb] blocks by one-third of lattice along the diagonal line, which was not reported in TMDCs either. Moreover, abrupt lattice contractions emerge along the $c$-axis at the transitions on warming, the magnitude of which is comparable to that of systems showing gigantic negative thermal expansion (NTE). The thermodynamic stability and the origin of the transitions are discussed from first-principle calculations. At low temperature, Pb(Ta$_{1+x}$Se$_{2}$)$_{2}$ exhibits a superconducting phase transition with $T_\textrm{c}\approx$ 1.8 K. From symmetry analysis and calculations, all three phases host topological nodal lines in the band structure. All these findings expand our understanding of polymorphic transitions and would be appealing to the field of intercalation chemistry, stack-engineering and quantum phenomena manipulation in TMDCs and beyond.~\cite{LiWB2021}.

\section{Results}

\subsection{Controlled phase selection} 
As shown in \textbf{Figure~1}, we propose a rational structural design in search of new TMDC-based compound, Pb(Ta$_{1+x}$Se$_2$)$_2$, taking inspiration from the misfit layered chalcogenide compounds (MLCs) with a wide variety of constituent ratios and layering schemes\cite{rouxel1995chalcogenide,Merrill2015}. 
Pb, Ta and Se powders with transport agent PbBr$_2$ were vacuum-sealed in a quartz tube and heated in a horizontal tube furnace with the temperature gradient (the details see in sample preparation section), which has been learned from and optimized in previous literature/experiments\cite{ShuangJiaPRB,yang2018superconductivity,yang2019enhanced,xhyang2021}. By varying the ratio of the atoms and the temperature gradient, we can selectively target 112-phase Pb-TaSe$_{2}$, PbSe-TaSe$_{2}$ \cite{yang2018superconductivity}, PbSe-(TaSe$_{2}$)$_{2}$ \cite{yang2019enhanced} or 124-phase Pb-(TaSe$_{2}$)$_{2}$ as the major product, among which the two MLCs are composed of alternating rocksalt-type PbSe layers and TaSe$_2$ layers along the $c$-axis. 

Noticeably, the 124-phase displays regular triangle shape, in contrast to the irregular/regular hexagonal shapes for MLCs/112-phase. More intercalated phases could be explored and discovered by varying the growth conditions including the atomic ratios, growth temperature and temperature gradient to cover more spaces of the experimental parameters.
\begin{figure*}[htb]
 \includegraphics[width=0.8\linewidth]{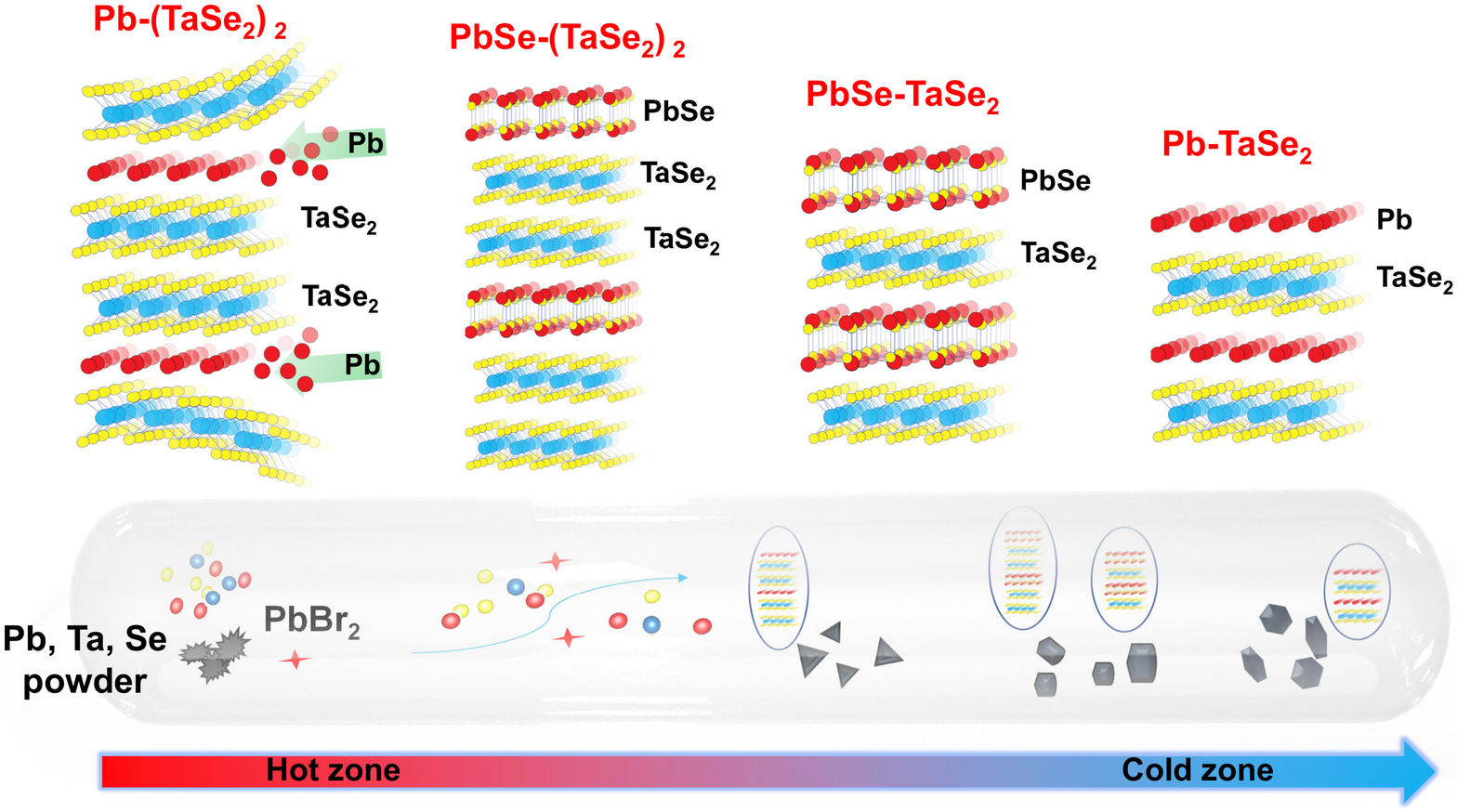}
 \Centering
  \caption{Schematic of Pb-Ta-Se related single-crystals grown by  chemical vapor transport (CVT). 
 Pb, Ta, and Se powders are taken by PbBr$_2$ agent from the source zone to the deposition zone, driven by the temperature gradient. The misfit compounds PbSe-TaSe$_2$ and PbSe-(TaSe$_2)_2$ display irregular hexagonal shape. 112-phase Pb-TaSe$_2$ and 124-phase Pb-(TaSe$_2)_2$ crystallize in regular hexagonal and triangle shape, respectively.}.
  \label{Fig1}
\end{figure*}

\subsection{Basic characterizations}
\textbf{Figure~2}a,b present high-angle annular dark-field high-resolution scanning transmission electron microscopy (HAADF-HRSTEM) images along [100] and [120] zone axes for the cross section at room-$T$ for Pb(Ta$_{1+x}$Se$_2$)$_2$ single-crystal. In comparison with the lattice structure shown in Figure~2c, the stacking sequence is clearly resolved. Pb(Ta$_{1+x}$Se$_2$)$_2$ can be viewed as stage-2~\cite{dresselhaus1981intercalation} Pb intercalation of 3$R$-TaSe$_2$, in which each two TaSe$_2$ slabs intercalated by one Pb layer compose the unit of building blocks. As seen in the Figure~2c, one unit cell (UC) of Pb(Ta$_{1+x}$Se$_2$)$_2$ is formed by the stacking of three building units in a special sequence. As a result, Pb(Ta$_{1+x}$Se$_2$)$_2$ crystallizes in a non-centrosymmetric trigonal lattice with space group (SG) $R3m$ (No. 160) (see below for detailed discussions). On close inspection of the HAADF-HRSTEM images, we observe additional Ta atoms randomly distributed in-between two TaSe$_2$ slabs, as marked by red arrows. Note that the existence of additional Ta atoms was also observed in 3$R$-Ta$_{1+x}$Se$_2$~\cite{Bjerkelund1967,Tanaka2020}, the amount of which is denoted by 2$x$ in Pb(Ta$_{1+x}$Se$_2$)$_2$. The selected area electron diffraction (SAED) patterns are presented in the inset, which allows for the clear observation of (00$\underline{3\ell}$) reflections and the determination of periodicity of 46.2 $\textrm{\AA}$ along the $c$-axis. 
\begin{figure*}[htb]
 \includegraphics[width=0.95\linewidth]{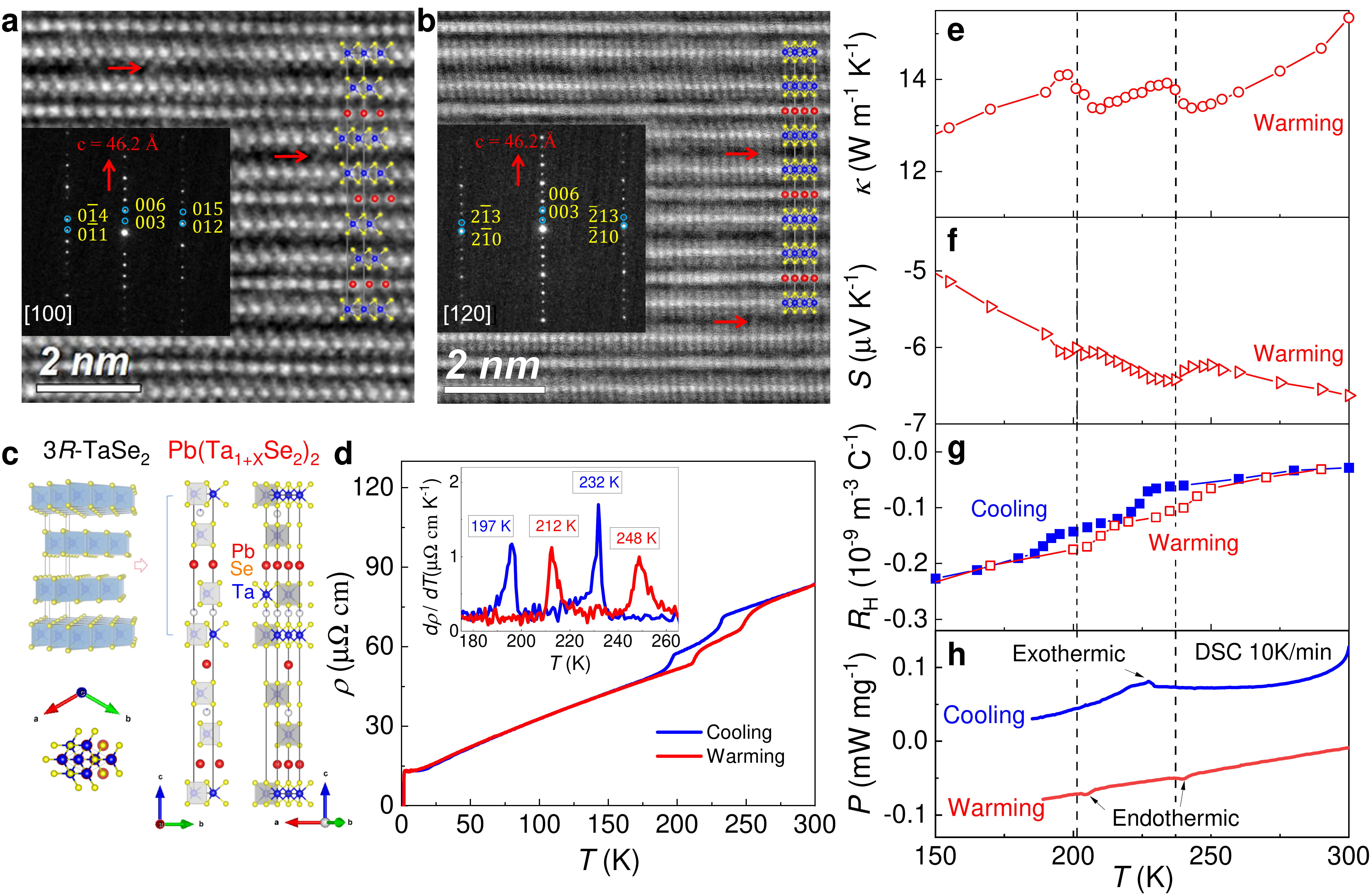}
 \Centering
  \caption{Basic properties of Pb(Ta$_{1+x}$Se$_2$)$_2$ single-crystals. a,b) HAADF-HRSTEM images along [100] and [120] zone axes. The red arrows mark interstitial Ta atoms. The insets show the corresponding SAED patterns. c) Crystal structure of $3R$-TaSe$_2$ (left) and unit cell of Pb(Ta$_{1+x}$Se$_2$)$_2$, as viewed in planes normal to [100] and [120] axes, respectively (right). d) $T$-dependent electrical resistivity on cooling and warming, which shows two step-like anomalies. The inset shows the $T$-derivative of resistivity, displaying the corresponding transition peaks. e-g) $T$-dependent thermal conductivity ($\kappa$), Seebeck coefficient ($S$) and Hall coefficient ($R_\textrm{H}$). h) DSC measurements on warming and cooling. The transition dips (peak) are marked by arrows. Two vertical dashed lines mark two transition temperatures.}
  \label{Fig2}
\end{figure*}

Figure~2d presents the $T$-dependent resistivity between 1.5 K and 300 K. At about 230 K, more prominent features are observed. The resistivity exhibits two consecutive, step-like jumps with apparent thermal hysteresis. Determined from the peaks of the temperature-derivative data shown in the inset of Figure~2d, the characteristic temperatures of consecutive jumps amount to $T_1\approx248~\textrm{K}$ and $T_2\approx212~\textrm{K}$ on warming and $T_1\approx232~\textrm{K}$ and $T_2\approx197~\textrm{K}$ on cooling. The anomalies are also observed in other samples at similar temperatures (see Figure~S1a in Supporting Information for details). Such distinct hysteresis is not from experimental artifacts, but implies the first-order like nature of the transitions. Moreover, the consecutive transitions are also manifested by two jumps in other transport measurements, including thermal conductivity ($\kappa$), Seebeck coefficient ($S$) and Hall coefficient ($R_\textrm{H}$) in Figures~2e-g. The linear Hall resistivity as a function of magnetic field is presented in Figure~S1b (Supporting Information), the negative sign of which implies electrons dominate in the transport.

In Figure~2h, differential  scanning  calorimetry (DSC) curve shows two clear absorption transition dips, when measured on warming, and a release peak in the reverse cycle. The dips (peak) refer to the absorption (release) of latent heat, signifying first-order phase transitions in conformity with the results from transport measurements. Note that, when cooling, the second transition peak is just out of the range of measurements of our instrument.

What kind of transitions are they? We first rule out magnetic phase transitions, since the $T$-dependent magnetic susceptibility ($\chi$) does not show any anomalies as seen in Figure~S2~(Supporting Information). On the other hand, we note that multi-step phase transitions are by no means rare in TMDCs~\cite{1T-TaS2}. For instance, 2$H$-TaSe$_2$ exhibits two-step CDW transitions from high-$T$ normal state to intermediate incommensurate state, and to low-$T$ commensurate state~\cite{Craven1977,Moncton1977,Freitas2016,Vescoli1998}. Similar results were also observed in In$_x$TaSe$_2$~\cite{Xu2021}. CDW transitions usually open up a band gap and suppress the density of states near the Fermi level, leading to remarkable changes in Hall signals. It would be tantalizing to attribute the consecutive transitions to CDW, since $R_\textrm{H}$ of Pb(Ta$_{1+x}$Se$_2$)$_2$ varies with temperatures in Figure~2g.

\subsection{X-ray diffraction on the (001) plane} 
To gain more information, we preformed X-ray  diffraction (XRD) measurements on the (001) plane of Pb(Ta$_{1+x}$Se$_2$)$_2$ single-crystal. XRD patterns at room-$T$ are presented in \textbf{Figure~3}a, the peaks are fully indexed by (00$\underline{3\ell}$) and the extracted $c$-axis lattice constant amounts to 46.134 \AA, close to TEM value.

\begin{figure}[htbp]

\includegraphics[width=0.5\linewidth]{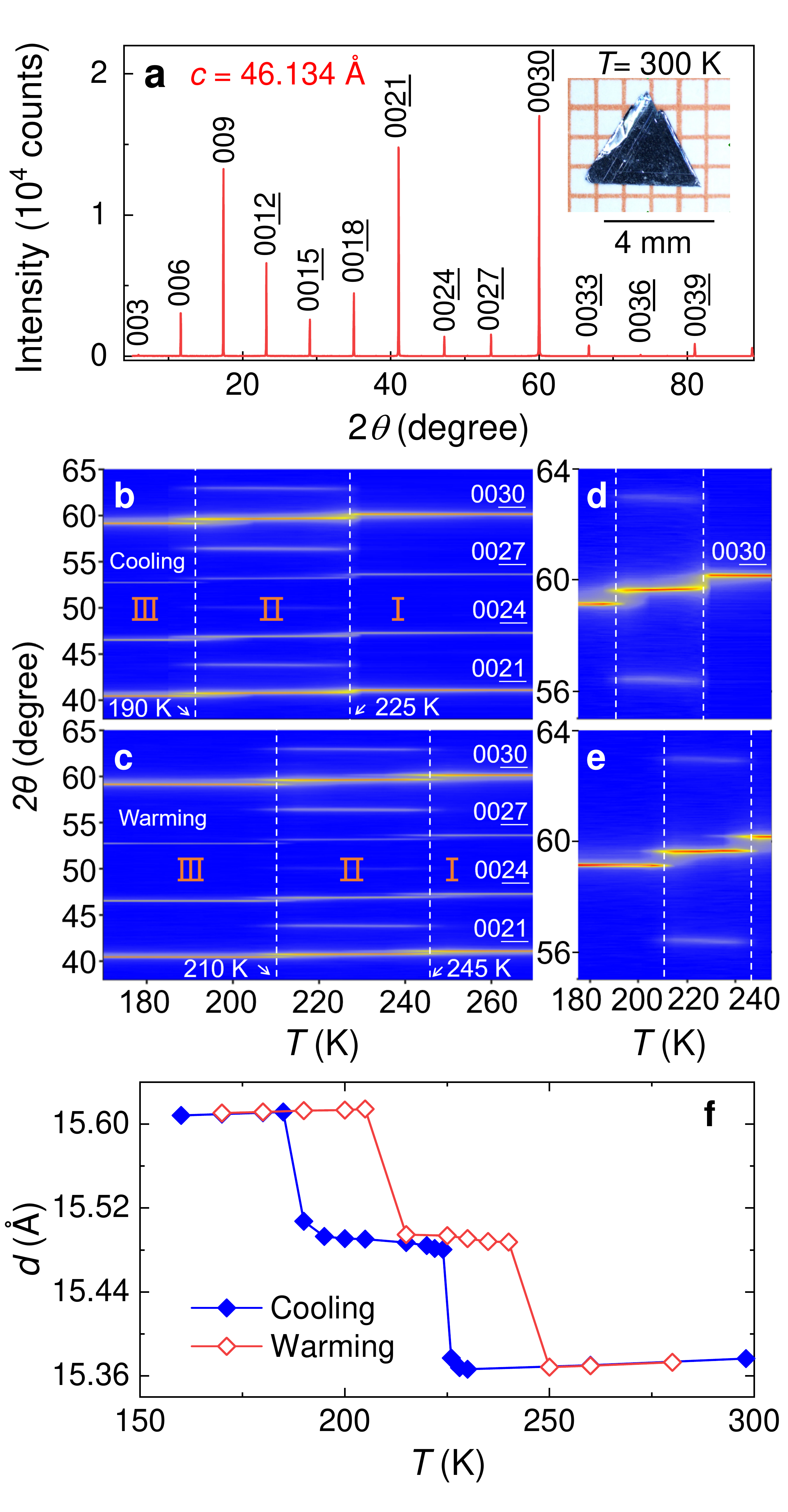}
\Centering
  \caption{X-ray diffraction on the (001) plane of Pb(Ta$_{1+x}$Se$_2$)$_2$ at various temperatures. a) XRD patterns at room-$T$. The inset shows the image of a millimeter-size specimen. b,c) Color maps of XRD patterns at various temperatures measured on cooling and warming, respectively. The high-$T$, intermediate and low-$T$ phases are marked by I, II and III, respectively. d,e) Zoom-in figures around the (00$\underline{30}$) peak. f) Interlayer distance as a function of temperature.}
  \label{Fig3}
\end{figure}
 
Figures~3b,c show the color maps of XRD patterns at various temperatures during a cooling and warming cycle, respectively, the XRD patterns of which are presented in Figure~S3 (Supporting Information). A set of four (00$\underline{3\ell}$) peaks emerge at $2\theta$ between $35^\textrm{o}$ and $65^\textrm{o}$ at high-$T$ (phase-I). By cooling, additional peaks emerge in the intermediate phase (phase-II), which seems to imply the formation of superstructure along the $c$-axis as seen in a CDW phase. While, by further cooling, the side peaks disappear in phase-III and the pattern returns to a set of four peaks bearing a marked resemblance to phase-I, which is, however, opposed to usual two-step CDW transitions.

The zoom-in figures around the intensive reflection (00\underline{30}) are presented in Figures~3d,e, from which several observations are made. First of all, the transitions from phase-I to phase-II and from phase-II to phase-III occur at $T_1\approx225$ K and $T_2\approx190$ K on cooling, respectively, which is manifested by the discontinuous jump of (00\underline{30}) peak to lower angles and the emergence/redisappearance of side peaks. Second, two adjacent phases coexist within a temperature window of 10 K around $T_1$ or $T_2$. Third, on warming, similar results are derived, but with the transitions shifted to slightly higher temperatures: $T_1\approx245$ K and  $T_2\approx210$ K, as seen in Figure~3c. The discontinuous jump, remarkable thermal hysteresis and phase coexistence provide a firm evidence supporting first-order structural phase transitions. Note that: the slight difference in $T_1$ and $T_2$ from XRD and transport measurements is due to experimental artifacts, because an experimental determination of the first-order phase transition temperature is sensitive to the concrete measuring processes.

The interlayer distance ($d$) at various temperatures extracted from Figure~S3 (Supporting Information) is presented in Figure~3f. Surprisingly, the lattice shows abrupt contractions along the $c$-axis at the transitions on warming. Quantitatively, $d$ shrinks by about $0.8\%$ at each step of transitions (totally $1.6\%$ from 200 K to 250 K). The magnitude of the thermal contractions is of the same order as that of systems showing gigantic NTE~\cite{Arvanitidis2003,Azuma2011,Takenaka2017}.

\subsection{Synchrotron single-crystal X-ray diffraction} 
At this point, the nature of the structural phase transitions remains fuzzy. Thus, we performed synchrotron single crystal XRD measurements on Pb(Ta$_{1+x}$Se$_2$)$_2$ at different temperatures, according to which the crystal structures of three phases are excellently resolved as shown in \textbf{Figure~4} and some key parameters are listed in \textbf{Table~1}. The detailed data collection, processing as well as resulting parameters including atomic coordinates, occupancies, atomic displacement parameters (ADP), etc., are included in Figures~S4-S5 (Supporting Information), Table~S1 and Supporting Information. 

Figure~4a presents the diffracted reflections in ($hk$0) planes at 298 K (phase-I), 230 K (phase-II) and 100 K (phase-III) (see Figure~S6 in Supporting Information for reflections in ($h0l$) planes). For phase-I and phase-III, the observed reflection condition is $-h+k+l=3n$, corresponding to the obverse setting of the $R$-lattice. The two phases are well resolved by SG $R3m$ with excellent figure of merits as shown in Table~1:  $R_\mathrm{obs} = 0.0226$ and $0.0205$ for phase-I and phase-III respectively. For improving the refinement, additional interstitial Ta atoms were introduced in-between two TaSe$_2$ layers (also seen in TEM data), the occupancy of which is $2x\approx$ 0.055(5) in Pb(Ta$_{1+x}$Se$_2$)$_2$. The details are included in the Supporting Information. 
\begin{figure*}[htpb]
\includegraphics[width=0.95\linewidth]{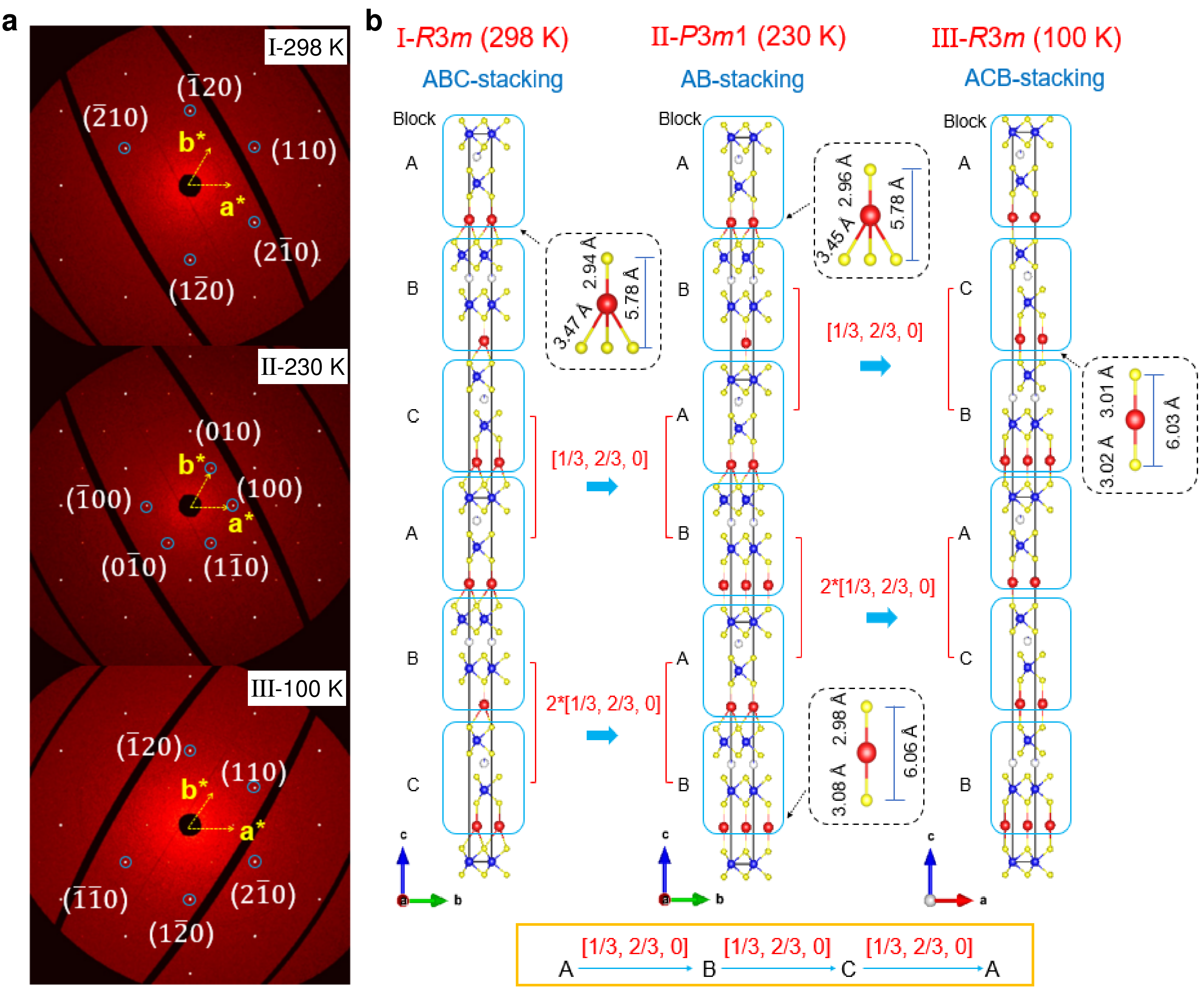}
\Centering
\caption{Synchrotron single crystal X-ray diffraction on Pb(Ta$_{1+x}$Se$_2$)$_2$ at different temperatures. a) Diffracted reflections in ($hk0$) planes at 298 K, 230 K and 100 K for phase-I, phase-II and phase-III, respectively. b) Crystal structures and Pb-Se coordinations of three phases derived from (a). Three phases are characterized by different stacking orders of building blocks (A, B, C). Note that phase-I and phase-II are viewed in a plane normal to the $a$-axis, but phase-III is viewed along the $b$-axis, which is consistent with the fact the diffracted reflections show a $60^\circ$ rotation along the $\mathbf{c}$* in phase-III as shown in (a).}
\label{Fig4}
\end{figure*}

On close inspection of the diffraction patterns, we find there is a $60^\circ$~rotation along the \textbf{c}* between phase-I and phase-III. The difference in the orientation matrices arises from distinct stacking orders that will be discussed in detail below. The absence of superlattice signals in phase-III in the $a$*$b$*-plane and along the \textbf{c}*-axis unambiguously rules out CDW transitions. This conclusion is further supported by low-energy electron diffraction (LEED) experiment as seen in Figure~S7~(Supporting Information).

\begin{table}[!htb]
\renewcommand\thetable{1}	
\Centering
  \caption{\label{tab1} Main structural refinement parameters in three phases for Pb(Ta$_{1+x}$Se$_2$)$_2$ at different temperatures ($T$). $a$, $b$ and $c$ denote lattice parameters and $\alpha$, $\beta$ and $\gamma$ are mutual angles. $Z$: the number of chemical formula units per UC. $V$: the UC volume. $R_\mathrm{int}$: the closeness of agreement in intensities for supposedly equivalent reflections. $R_\mathrm{obs}$: the figure of merit for the refinement result. $I$: the intensity of the reflection. $\sigma$: the standard uncertainty of a parameter.}

  \begin{tabular}{c|ccc}
  \hline
                    & Phase-I         & Phase-II        & Phase-III       \\
  \hline
  $T$ (K)                  & 298           & 230           & 100           \\
  SG   & $R3m$           & $P3m1$            & $R3m$             \\
                          & \multicolumn{3}{c}{ $a=b$; $\alpha = \beta = 90^\circ$, $\gamma=120^\circ$}   \\
  $a$ (\AA)                           & 3.43910(10) & 3.43710(10) &  3.43390(10) \\
  $c$ (\AA)                            & 46.1654(8)  & 31.0096(4)      & 46.8308(7)  \\
  $Z$                             & 3               & 2               & 3               \\
  $V/Z$ (\AA$^3$)                      & 157.63(2)       & 158.63(5)   & 159.41(2)    \\

  $R_\mathrm{int}$                &0.0422               &0.0395       & 0.0451 \\

  $R_\mathrm{obs}$($I>3\sigma(I)$)  &    0.0226     &  0.0676             & 0.0205 \\
  \hline
  \end{tabular}
\end{table}

The diffraction pattern of phase-II at 230 K is a bit complex due to the partial overlapping of two adjacent phases for first-order phase transitions, which obstructs a direct identification of this intermediate phase. We propose a reasonable hypothesis and attribute phase-II to SG $P3m1$ (No.~156). $R_\mathrm{obs}$ value for the refinement, amounting to 0.0676, remains in a reasonable regime. The details supporting the hypothesis are presented in the Supporting Information. The lattice parameters of three phases are presented in Table~1. The material shows large contractions along the $c$-axis, the magnitude of which equals to that in Figure~3f within the experimental margin, but leaving the $ab$-plane almost intact from 100 K to 298 K. The gigantic, abrupt contractions could be broadened by introducing relaxors~\cite{Azuma2011,Takenaka2005}, such as doping, which may pave the way for potential applications of NTE in thin film TMDCs.

At this point, we achieve a firm conclusion that the two consecutive first-order phase transitions are attributed to structural phase transitions from high-$T$ $R3m$ phase to intermediate $P3m1$ phase and back to low-$T$ $R3m$ phase. More intriguingly, $R3m$ and $P3m1$ phases share the same point group symmetry, indicating that the apparent signals we observed in the transport and XRD measurements are merely related to subtle changes of the lattice.

In Figure~4b, we present the crystal structures of three phases, all of which contain the same building block [Se-Ta-Se-Ta$_{2x}$-Se-Ta-Se-Pb], abbreviated by [Pb(Ta$_{1+x}$Se$_2$)$_2$]. For phase-I, the lattice orders in a special sequence ABCABC... due to the $R$-lattice centering in each UC. Here, block B is translated from block A by a vector $\textbf{T}=$ [1/3, 2/3, 0] and block C translated from block B by the same vector. For the transition from phase-I to phase-II, two UCs, consisting of ABCABC stacking, is transformed into three UCs of ABABAB stacking, wherein each pair of blocks slide rigidly by \textbf{T} with respect to its neighbor on top of it, i.e. blocks CA slide by \textbf{T}, the neighboring BC slide by additional \textbf{T}, according to which the next blocks AB slide by another additional \textbf{T} resulting in a lattice translation (3\textbf{T} = [1, 2, 0], equivalent to no slide). The second transition from phase-II to phase-III involves the transformation of ABABAB stacking to ACBACB stacking, in which each pair of blocks translate in the same manner as that of the first transition, i.e. blocks BA  slide by \textbf{T} and the neighboring BA slide by additional \textbf{T}.

In phase-I, the coordination geometry of Pb is tetrahedral with a short vertical Pb-Se link and three long leaning Pb-Se links, as seen in Figure~4b. The length of the short link is within the range of Pb-Se bonding distance (3.1~\AA)~\cite{PbSe_Saito}, while the long links (3.5~\AA) is far out of range implying almost no or negligible Pb-Se bonding. When the blocks slide at the transitions, the long links break, leading to reconfiguration of Pb-Se tetrahedrons. Consequently, linear dumbbell-like Se-Pb-Se (PbSe$_2$) bonds, with short Pb-Se distance (3.01~\AA), form in phase-III. In the intermediate phase-II, both tetrahedral and dumbbell-like Pb-Se coordinations exist at the same weight. The vertical thickness of PbSe$_4$ tetrahedrons is smaller than that of PbSe$_2$ dumbbells, which accounts for the anisotropic lattice contractions at transitions on warming.

In short, we have resolved that the two-step anomalies in Pb(Ta$_{1+x}$Se$_2$)$_2$ arises from unique, consecutive, first-order structural phase transitions, associated with the stacking degree of freedom, i.e., the commensurate slide of the building blocks along the diagonal line. Our results may give a hint to the nature of the first-order structural phase transition in PbTaSe$_{2}$ 112-system~\cite{XuXF2017,Sergey2017}. To the best of our knowledge, such kind of transitions has not been reported in TMDCs as well as other van der Waals electronic materials. A search of literature only reveals that Na$_x$CoO$_2$, as a cathode of sodium-ion batteries, seems to exhibit a similar, but only one-step transition by penetration/extraction of Na-ions during the charging/discharging cycles~\cite{Wang2017NaxCoO2}.

\subsection{Superconducting properties} 
\textbf{Figure~5}a presents the $T$-dependent resistivity at low temperatures, from which Pb(Ta$_{1+x}$Se$_2$)$_2$ exhibits a sharp superconducting transition with an onset $T_\textrm{c}\approx$ 1.8 K, determined at the 90\% of the normal state value. The superconductivity (SC) is also revealed by direct current (DC)  magnetic susceptibility in the inset. The SC shielding volume fraction is close to 100\% from the zero-field cooling (ZFC) process.

\begin{figure}[hbtp]
 \includegraphics[width=0.5\linewidth]{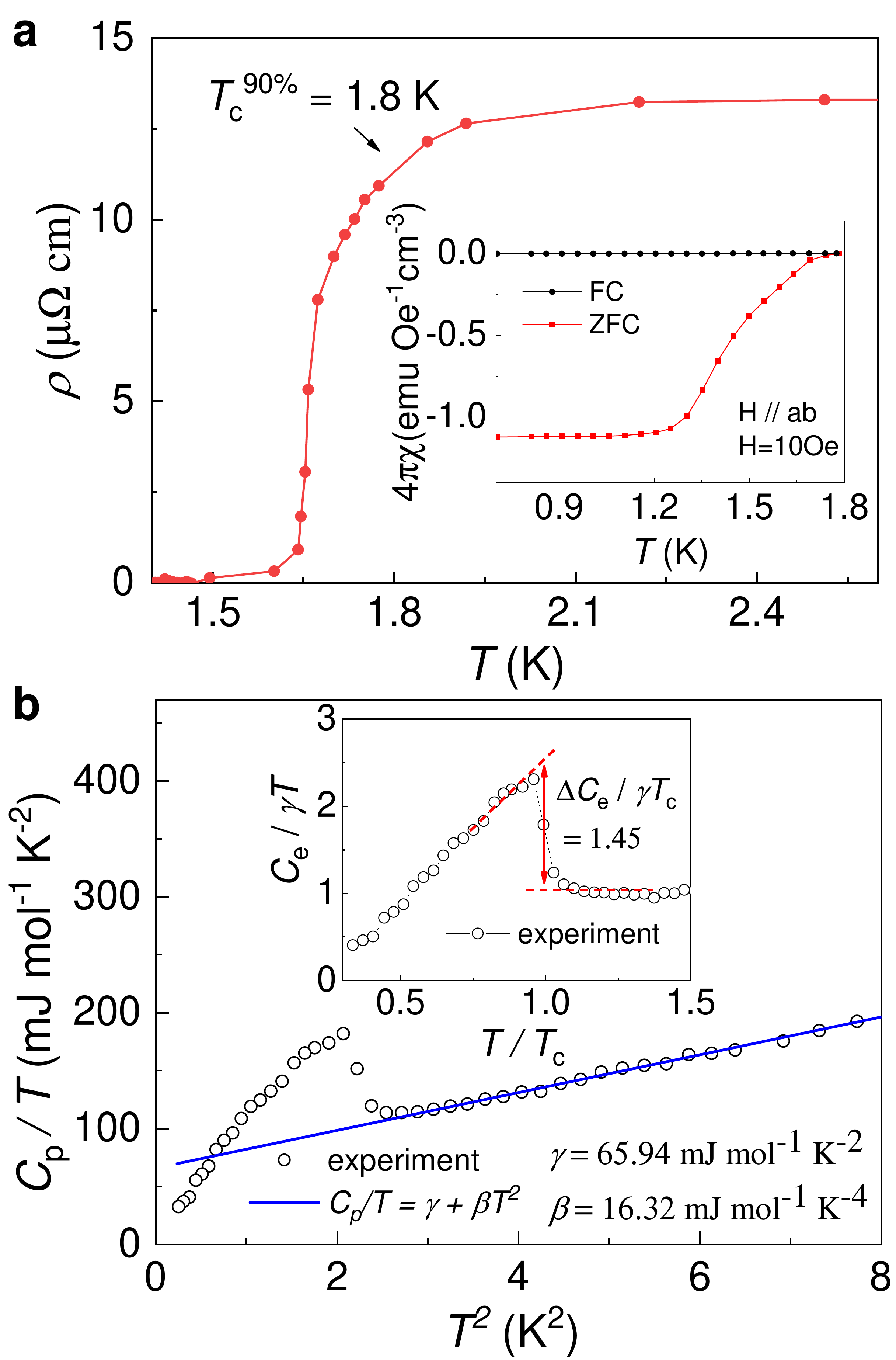}
 \Centering
  \caption{Superconductivity of Pb(Ta$_{1+x}$Se$_2$)$_2$.  a) $T$-dependence of resistivity around $T_\textrm{c}$. Inset: $T$-dependence of DC magnetic susceptibility at $H$//$ab$ and $H$ = 10 Oe on zero-field cooling (ZFC) and field cooling (FC), respectively. b) Specific heat ($C_\textrm{p}$) presented in the form of $C_\textrm{p}/T$ versus $T^2$.
The blue line is a linear fit to the expression $C_\textrm{p}/T=\gamma+\beta T^2$. 
Inset: normalized electronic specific heat $C_\textrm{e}/\gamma T$ versus $T/T_\textrm{c}$. The red dashed lines and arrow mark the jump at the transition.} 
  \label{Fig5}
\end{figure}

The bulk SC is further confirmed by the remarkable specific heat ($C_\textrm{p}$) jump around $T_\textrm{c}$ as seen in the Figure~5b. By fitting $C_\textrm{p}$ at normal state to the relation
of $C_\textrm{p}/T = \gamma+\beta T^2$, one can yield the Sommerfeld coefficient $\gamma\approx$ 65.94 mJ~mol$^{-1}$~K$^{-2}$ and the lattice coefficient $\beta\approx$ 16.32
mJ~mol$^{-1}$~K$^{-4}$. According to  the relation $\Theta_\textrm{D} = (12\pi^4NR/5\beta)^{1/3}$, the Debye temperature ($\Theta_\textrm{D}$) is estimated to be 136 K, wherein $N$ is the number of atoms per
formula unit and $R$ is the gas constant. By subtracting the lattice contribution, the electronic part ($C_\textrm{e}$) is presented in the inset. The normalized jump $\Delta C_\textrm{e}/\gamma T_\textrm{c}$ at the transition amounts to 1.45, which agrees well with the weakly coupled BCS value 1.43\cite{BCS}.

\section{Discussion}

\noindent We have achieved the main results of this work, whereas the underlying physics remains to be addressed. Thus, we performed first-principle calculations. The difference in the internal energy ($\triangle U_\textrm{in}$) calculated at zero-$T$ is merely a few tens of meV per unit block between adjacent phases: $\triangle U_\textrm{in(I-II)}\approx27.3$ meV and $\triangle U_\textrm{in(II-III)}\approx27.1$ meV, which implies the thermodynamic stability of phase-III at low-$T$ (see the method in the Supporting Information). On the other hand, as shown in Table~S1, Pb atoms in tetrahedral Se voids show much larger ADP than that of other atoms in the same structure at high-$T$, while ADP of Pb in a dumbbell geometry is only slightly larger than that of other atoms at low-$T$. The much larger ADP of Pb atoms at high-$T$ may provide a large vibrational entropy ($S_\textrm{vb}$) due to the extended phase space in terms of moment and position~\cite{vibration_entropy_flutz}. We argue that small $\triangle U_\textrm{in}$ could be compensated by large $\triangle S_\textrm{vb}$, leading to lower Helmholtz free energy ($F=U_\textrm{in}-TS_\textrm{vb}$) at higher-$T$, which provides a thermodynamic basis for the structural transitions~\cite{vibration_entropy,vibration_entropy_tony}.

 \begin{figure*}[hbt]
\includegraphics[width=0.95\linewidth]{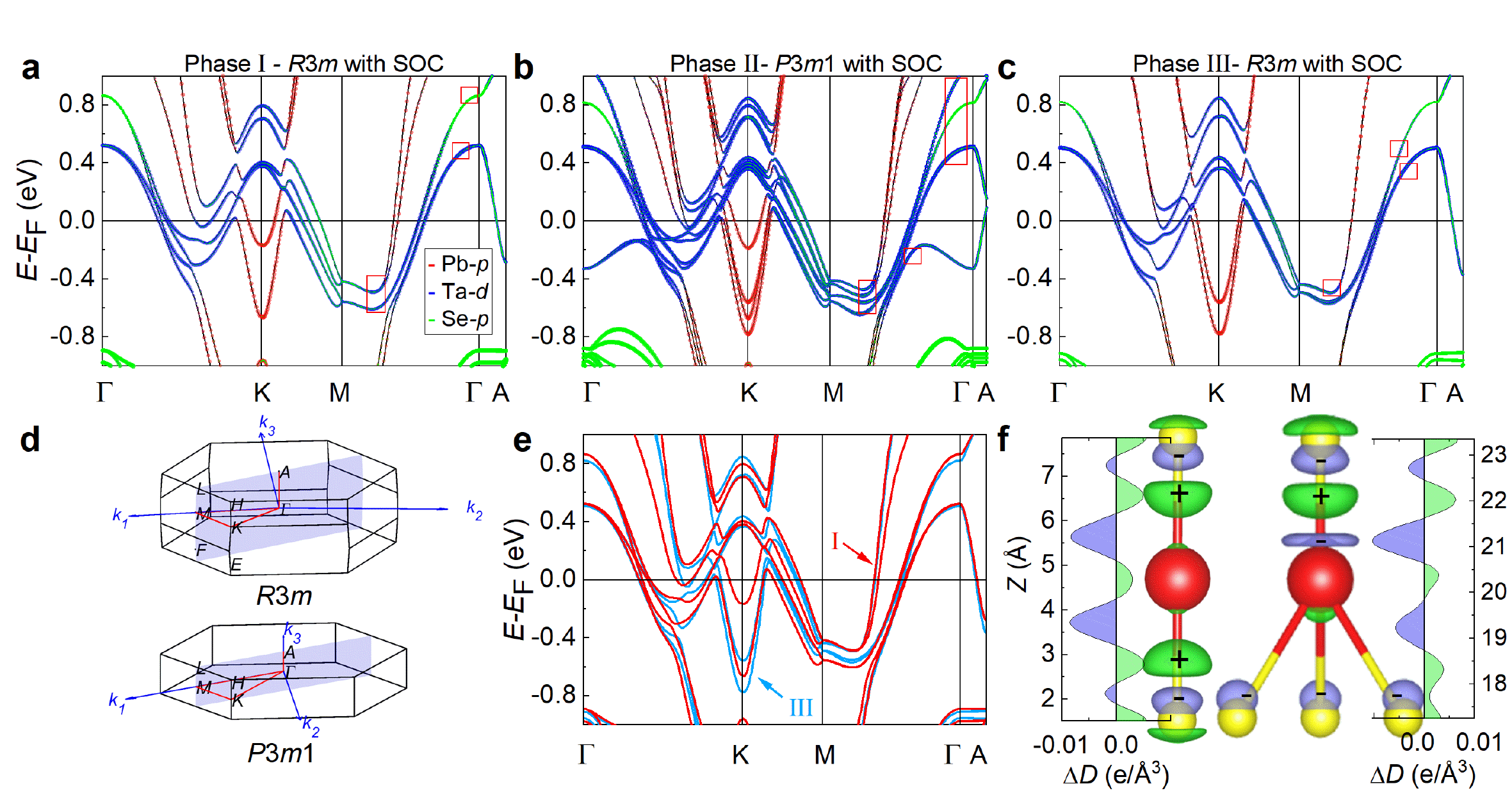}
\Centering
  \caption{Calculated band structures of Pb(Ta$_{1+x}$Se$_2$)$_2$ with SOC for three phases. a-c) Band structures close to the Fermi level for three phases. The contributions from Pb-$6p$, Ta-$5d$ and Se-$4p$ orbitals are colored in red, blue and green, respectively. The red boxes enclose topological line-nodes along the $M$--$\varGamma$ high symmetry path. d) First Brillouin zones with high symmetry points for SG $R3m$ and $P3m1$. The red lines mark the high-symmetry path as shown in (a-c). The purple plane is one of the vertical mirror plane where the nodal lines reside. e) Overlapping band image for phase-I (red) and phase-III (cyan). f) Planar-averaged charge density difference ($\Delta D(z)$) and charge displacement in real space for two Pb-Se coordinations: dumbbell (left) and tetrahedron (right). The isosurface level is set as 0.015 electron/\AA$^3$. Charge accumulation and depletion are marked in green and purple, respectively.}
  \label{Fig6}
\end{figure*}

Electronically, the calculated band structure with spin-orbital coupling (SOC) along high-symmetry paths are presented in \textbf{Figure~6}a-c for three phases. In Figure~6b, phase-II exhibits multi-fold bands because of the reduced volume of Brillouin zone in SG $P3m1$, in comparison with SG $R3m$ as seen in Figure~6d. According to the irreducible representation analysis, we find Pb(Ta$_{1+x}$Se$_2$)$_2$ hosts topological nodal-lines, marked by red boxes in Figures~6a-c, in three phases. The detailed calculations, including the temperature evolution of nodal-line morphology, are presented in Figures~S8-S10 and the Supporting Information.

Since phase-I and phase-III host the same SG, we can compare their band structure directly in Figure~6e. At $K$-point, two conducting bands, mainly composed of Pb-6$p$ orbitals, shift by about 0.13 eV and 0.40 eV respectively, to lower electron binding energies from phase-III to phase-I, implying the electron loss of Pb. 
Other bands around $M$ point and along $\varGamma$-$K$ line, mainly composed of Ta-5$d$ and Se-4$p$ orbitals, show much slighter shift due to the large density of state (DOS) near the Fermi level.

Figure 6f presents the planar-averaged electron density difference in real space for two kinds of Pb-Se coordination (see the method in Figure~S11 and the Supporting Information). For PbSe$_2$ dumbbells, electron density accumulates in-between the Pb-Se link, indicating strong hybridization of Pb-6$p$ and Se-4$p$ orbitals. While for PbSe$_4$ tetrahedrons, the vertical link shows hybridization alike PbSe$_2$ dumbbells, however, the three leaning links do not show remarkable charge accumulation, implying weak Pb-Se coupling. All these are in conformity with the aforementioned discussions that the short vertical links favor strong chemical bonding and the long leaning links are characterized by negligible bonding. It also provides a well explanation for the large ADP of PbSe$_4$ tetrahedrons. 

Overall, we conclude that the charge transfer between Pb and TaSe$_2$ layers underlies the unique stacking phase transitions. The charge transfer is also manifested by the slight shift of Pb-5$d$ shallow core levels in the photoelectron spectra, as seen in Figure~S12 (Supporting Information). It also explains the temperature variation of $R_\textrm{H}$ as shown in Figure~2g.

Regarding the fact that the successful synthesis of pure 3$R$-TaSe$_{2}$ single-crystals is extremely challenging and only thin films can be obtained by advanced techniques such as chemical vapor deposition (CVD)~\cite{Zhengjacs} or molecular-beam epitaxy (MBE)~\cite{Tanaka2020}, the stabilization of rhombohedral stacking of TaSe$_{2}$ layers by Pb intercalation in a bulk form by a conventional CVT method is quite appealing. In Pb(Ta$_{1+x}$Se$_2$)$_2$, CDW phase has been completely suppressed, in stark contrast with the pristine 3$R$-TaSe$_{2}$, but the superconducting critical temperatures are close in these two phases. Since Pb(TaSe$_2$)$_n$ with $n=$ 1 and 2 has been successfully synthesized, it would be promising to explore higher order phases by varying the growth condition. If this is proved to be true, this system can act as an ideal platform to study continuous intercalation effect in TMDCs.

\section{Conclusion}
In summary, this work synthesized the first specimen of 124-phase of metal intercalated TMDCs. 
The prospect of growing higher order Pb(TaSe$_2$)$_n$ phases makes this system an ideal template to study metal intercalation chemistry of TMDCs. The 124-phase Pb(Ta$_{1+x}$Se$_2$)$_2$ exhibits nontrivial, two-step, reversible, first-order structural phase transitions featured by rigid slide of building blocks by one-third of the lattice along the diagonal line in the $ab$-plane. The commensurate transitions related to stacking degree of freedom result in unique stacking orders: ABC, AB and ACB, which will certainly enrich our understanding of polymorphic phase transitions in TMDCs and beyond. The interpretation of the underlying physics may shed light on the realization of stacking sequence engineering in TMDCs. 

We suggest that a further study to broaden the lattice contractions in Pb(Ta$_{1+x}$Se$_2$)$_2$, combined with its lattice compatibility with other TMDCs, may open a possibility for potential applications of gigantic thin film NTE effects. The study of the relation between superconductivity, band topology, stacking transitions in this system would also be of great interest in future work.

\section{Experimental Section}

\textit{Sample preparation:}\quad Single crystals of Pb(Ta$_{1+x}$Se$_2$)$_2$ were prepared by CVT method from a mixture of high-purity Pb, Ta, Se powders in an appropriate ratio $1:2:4.05$, in which additional PbBr$_2$ (10 mg/cm$^3$ in concentration) was used as transport agent. The powders were thoroughly mixed and sealed in an evaluated quartz tube with a diameter of 12 mm and a length of 15 cm. Then, the tube was placed in a horizontal two-zone furnace for a week by setting the temperature at 900~$^\textrm{o}$C and 800~$^\textrm{o}$C (powders at the hot side), respectively. The single crystals were found in the middle of the tube with a typical dimension of 3$\times$3$\times$0.2 mm$^{3}$.

\textit{Physical property measurements:}\quad $T$-dependent resistivity was measured in an Oxford superconducting magnet system equipped with a $^3$He cryostat. $T$-dependent thermal conductivity, thermopower and Hall resistivity were measured in a Quantum Design physical property measurement system (PPMS). DC magnetization was measured in a Quantum Design magnetic property measurement system (MPMS3) equipped with a $^3$He cryostat. DSC measurements were performed in a Metter Toledo DSC 3+ system at a heating/cooling rate of 10 K/min.

The HAADF-HRSTEM images and the SAED patterns were obtained in a state-of-the-art STEM with probe Cs corrector (Titan G2 80-200 ChemiSTEM, FEI Co., Hillsboro, OR) operated at 200 kV. The core-level photoelectron spectra were measured by using monochromatic He II ($hv=$ 40.82 eV) radiation at normal emission. LEED measurements were performed by a BDL800IR spectrometer with beam energy of 100 eV.

The powder XRD on the (001) plane of a single-crystal was carried out by using a PANalytical X-ray diffractometer (Model EMPYREAN) with monochromatic Cu-K$_{\alpha1}$ radiation. $T$-dependent data was collected under an atmosphere of $^4$He in a sealed low-temperature stage.  Synchrotron single crystal XRD measurements were performed at station EH1 of beamline P24 of PETRA-III at DESY. The full details of data collection and processing are summarized in the Supporting Information.

\textit{Band structure calculations:}\quad Density functional theory (DFT) calculations were carried out by using the Vienna \textit{ab initio} simulation package ({\sc vasp})~\cite{PhysRevB.54.11169}, based on the generalized gradient approximation (GGA) method under the Perdew-Burke-Ernzerhoff (PBE) parameterization~\cite{PhysRevLett.77.3865}. The irreducible representations of electronic eigenstates at different \textit{k}-points are determined by the software package \texttt{irvsp}~\cite{gao2021irvsp}. Wannier functions are constructed by projecting Bloch states onto Pb-$6p$, Ta-$5d$ and Se-$4p$ orbitals through {\sc wannier90}~\cite{PhysRevB.56.12847, PhysRevB.65.035109, pizzi2020wannier90}. Nodal lines are computed with the {\sc wanniertools} package \cite{wu2018wanniertools} and the nodal points protected by symmetry are identified by \texttt{irvsp}~\cite{gao2021irvsp}. The details are included in the Supporting Information. 

\setlength{\parindent}{0em}
\section*{Supporting Information}  
Supporting Information is available from the Wiley Online Library or from the author.\\
~\\
CCDC deposition numbers 2111695, 2111700 and 2111702 contain the supplementary crystallographic data (CIF) for phase-I, II and III, respectively, in this paper.\\
~\\
All code supporting DFT calculations is available at \href{https://github.com/shan-ping/PbTa2Se4}{https://github.com/shan-ping/PbTa2Se4}.

\section*{Acknowledgements} 

\noindent We are grateful to Ruihua He, Shi Liu, Jianhui Dai, Toms Rekis and Quansheng Wu for helpful discussions. This research was supported by National Key Projects for Research \& Development of China (Grant No. 2019YFA0308602); National Natural Science Foundation of China via Project 11904294, 11774305 and 62004172;  Zhejiang Provincial Natural Science Foundation of China under Grant No. LQ19A040005 and the foundation of Westlake  Multidisciplinary Research Initiative Center (MRIC)(Grant No. MRIC20200402 and 20200101). J.-K. Bao acknowledges Alexander von Humboldt Foundation for financial support in Germany. We thank the support provided by Chao Zhang and Ying Nie from Instrumentation and Service Center for Physical Sciences (ISCPS) at Westlake University. Single-crystal X-ray diffraction experiments with synchrotron radiation source has been preformed at the beamline P24 of PETRA-III at DESY in Hamburg, Germany.

\section*{Conflict of Interest} 

The authors declare no conflict of interest.

\section*{Author Contribution} 
X.Y., J.-K.B. and Z.L. contributed equally to this work. 
X.Y. designed the experiment and grew the single-crystals with the help of C.J.. She was assisted in the measurements by J.W.. Y.L. performed powder XRD measurements under the supervision of G.-H.C.. P.L. performed the core level photo-emission measurements under the supervision of Y.L..
T.S. obtained the atomic-resolution HAADF-HRSTEM images. W.G. performed LEED measurements under the supervision of Y.N.. 
J.-K.B. performed the synchrotron single crystal XRD measurements and analyzed the data under the supervision of S.V.S. and with the help from S.R., S.R.K., M.T. and C.P.. 
Z.L. carried out first-principle calculations with the help from W.L..
X.Y., Z.L., J.-K.B. and X.L. wrote the manuscript. X.L. and Z-A.X. led the project. All authors contributed to the discussion. 

\section*{Data Availability Statement} 

The data that support the findings ofthis study are available from the corresponding author upon reasonable request.




\vspace{2em}

\begin{figure}[htb]
\textbf{Table of Contents}\\
\medskip
\includegraphics[width=5.5cm]{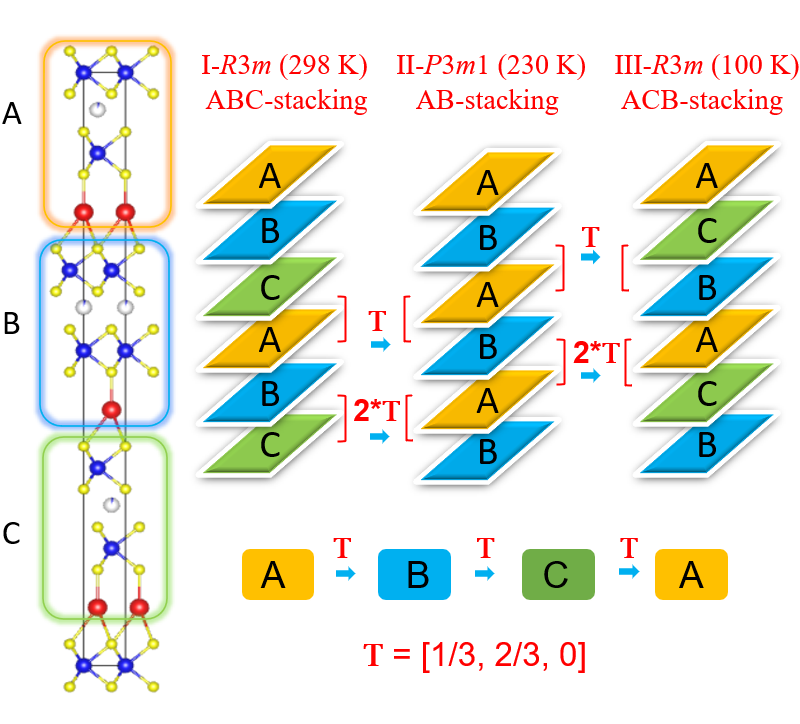}
  \medskip
  \caption*{The  transitions  are  solely  associated  with  the  stacking  degree  of  freedom,  evolving  from  a  high temperature phase with ABC stacking and symmetry  $R3m$ to an intermediate phase with AB stacking and $P3m1$, and finally to a low temperature phase with again symmetry  $R3m$, but with ACB stacking.  Each step involves a rigid slide of building blocks by avector [1/3, 2/3, 0].}
\end{figure}

\end{document}